\documentclass[journal]{vgtc} 
\usepackage{algorithm}
\usepackage{array}
\usepackage[caption=false,font=normalsize,labelfont=sf,textfont=sf]{subfig}
\usepackage{textcomp}
\usepackage{stfloats}
\usepackage{url}
\usepackage{verbatim}
\usepackage{graphicx}
\usepackage{cite}
\usepackage{color}
\usepackage{pifont}
\usepackage{amsmath,amsthm,amssymb,amsfonts}

\usepackage{algpseudocode}
\usepackage{appendix}
\graphicspath{{figures/}{pictures/}{images/}{./}} 

\usepackage{tabu}                      
\usepackage{booktabs}                  
\usepackage{lipsum}                    
\usepackage{mwe}                       

\usepackage{mathptmx}                  
\onlineid{1890}

\vgtccategory{Research}

\vgtcpapertype{please specify}

\title{TeamPortal: Exploring Virtual Reality Collaboration Through Shared and Manipulating Parallel Views}

\author{%
  \authororcid{Xian Wang}{0000-0003-1023-636X},
  \authororcid{Luyao Shen}{0009-0006-8403-6271}, 
  \authororcid{Lei Chen}{0000-0002-1524-3000}, 
  \authororcid{Mingming Fan}{0000-0002-0356-4712} and 
  \authororcid{Lik-Hang Lee}{0000-0003-1361-1612}
}

\authorfooter{
  \item
  	Xian Wang is with the Hong Kong Polytechnic University.
  	E-mail: xiann.wang@connect.polyu.hk.
  \item
  	Luyao Shen is with the Hong Kong University of Science and Technology (Guangzhou).
  	E-mail: lshen595@connect.hkust-gz.edu.cn.

  \item 
        Lei Chen is with Hebei GEO University.
  	E-mail: chenlei@hgu.edu.cn.

    \item 
        Mingming Fan is with the Hong Kong University of Science and Technology (Guangzhou).
  	E-mail: mingmingfan@ust.hk.
    
    \item 
        Lik-Hang Lee is with the Hong Kong Polytechnic University.
  	E-mail: lik-hang.lee@polyu.edu.hk (corresponding author).
    
}

\abstract{
Virtual Reality (VR) offers a unique collaborative experience, with parallel views playing a pivotal role in Collaborative Virtual Environments by supporting the transfer and delivery of items. Sharing and manipulating partners' views provides users with a broader perspective that helps them identify the targets and partner actions. We proposed TeamPortal accordingly and conducted two user studies with 72 participants (36 pairs) to investigate the potential benefits of interactive, shared perspectives in VR collaboration. Our first study compared ShaView and TeamPortal against a baseline in a collaborative task that encompassed a series of searching and manipulation tasks. The results show that TeamPortal significantly reduced movement and increased collaborative efficiency and social presence in complex tasks. Following the results, the second study evaluated three variants: TeamPortal+, SnapTeamPortal+, and DropTeamPortal+. The results show that both SnapTeamPortal+ and DropTeamPortal+ improved task efficiency and willingness to further adopt these technologies, though SnapTeamPortal+ reduced co-presence. Based on the findings, we proposed three design implications to inform the development of future VR collaboration systems.
} 

\keywords{TeamPortal, Virtual Reality, Collaboration, Share Perspectives, Parallel Views}

\teaser{
  \centering
  \includegraphics[width=\linewidth]{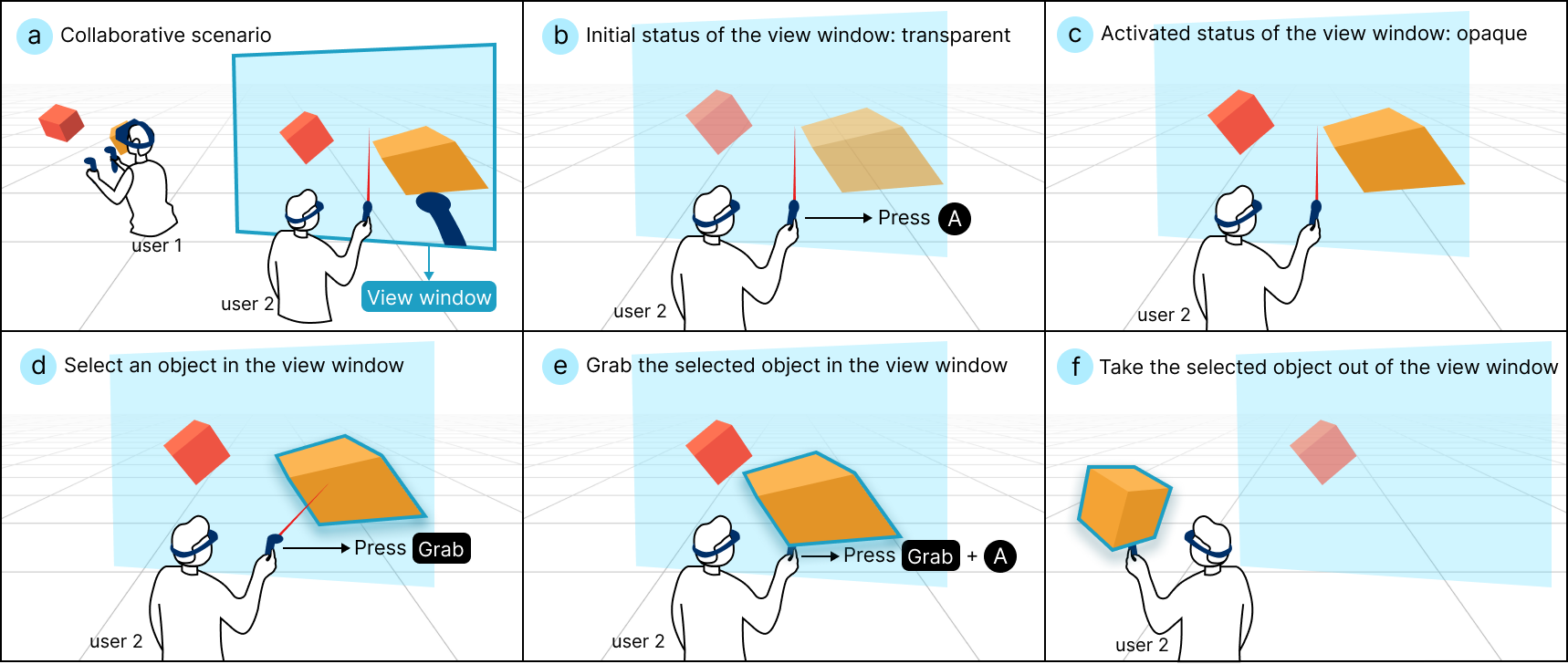}
  \caption{Illustration of TeamPortal interaction. TeamPortal allows users to share and manipulate each other's FOV. For example, User 2 can interact with User 1's FOV (example in this figure), and the same interaction applies in reverse. (a) Two users collaborate in a VR environment with TeamPortal where User 2's View Window synchronizes in real-time with User 1's perspective. (b) to (f) depict the step-by-step process of User 2 retrieving the orange cube from User 1's perspective through the View Window. The reverse process allows User 2 transfers a cube to User 1’s perspective by placing it into the View Window.}
  \label{fig:teaser}
}

\begin{document}



\maketitle

\section{Introduction}



VR technology is increasingly used in a variety of remote collaboration and socialization scenarios, enhancing people's ability to interact in different physical locations and changing the way they work, including virtual social venues~\cite{maloney2020talking,VRAlmostThere2023Sykownik,mei2021cakevr,freeman2022working}, virtual workplaces~\cite{olaosebikan2022identifying,schafer2019towards}, and industrial scenarios~\cite{henstrom2023immersive,gomez2024protocolvr,hoang2022evaluation}. Complex collaborative interactions are generally inevitable in multi-user VR environments, and the transfer and delivery of items between users are crucial, such as sharing documents~\cite{burova2022asynchronous,peng2021exploring}, delivering components~\cite{auda2021m,yildiz2019designing}, and distributing virtual media (e.g., content and information)~\cite{thanyadit2018efficient,li2019measuring}. 
In addition to these scenarios being discussed in controlled environments, more and more scenes in commercial multiplayer VR applications involve transferring virtual objects. For example, Rec Room~\cite{recroom} has VR rooms that can support multiplayer card and ball games. In such scenarios, cards and balls will be transferred between users. The \textit{Kitchen Cooks!} on VRChat~\cite{vrchat} is a kitchen that supports multiplayer cooking where users need to collaborate to fulfill orders. During collaboration tasks, ingredients are passed between users. In MeetinVR~\cite{meetinvr}, users can share sticky notes with other users in VR meeting rooms. 
However, in these virtual application scenarios, the focus of researchers and developers is entertainment or is limited to scenario-specific functionality. In contrast, collaborative interactions among users, such as the ability to transfer virtual instances to other users or retrieve virtual objects of interest from other users' locations, have not been extensively explored. 

The transfer of virtual items between users in VR collaboration may pose two major challenges. \textbf{(1) How to identify the virtual objects of interest to the users themselves and their partners.} In the real world, users may experience difficulty in quickly understanding the objects indicated by the partner due to the lack of the partner's field of view (FOV), i.e., either the FOV is blocked, or the object is outside the user's FOV~\cite{yu2019design,bork2018towards}. In virtual environments that alter the user's spatial judgment, this problem will be even more pronounced due to a narrower FOV than in the physical reality ~\cite{traquair1927introduction,grinyer2022effects,sauer2022assessment,banerjee2021side}. \textbf{(2) How the user can accurately and efficiently access virtual objects}~\cite{xia2018spacetime}. The user and the collaborator may be separated by a long distance in a virtual environment or even occlusion between them could deteriorate their collaboration, e.g., by a virtual structure or by the partner's avatar~\cite{parger2021unoc,yu2020fully}. Therefore, the user may not be able to directly reach the desired target object.

The sharing of FOV between collaborators could serve as a prominent solution to the aforementioned challenges in transferring virtual items. 
Employing a shared perspective that displays others' view could aid in discerning the positions and actions of their collaborators.
When users can know what their partner is looking at, it can improve understanding and coordination between users and enable more effective interactions~\cite{bovo2022cone}. Studies have shown that sharing one's FOV with partners can significantly improve the efficiency of collaborative tasks by providing contextual and visual information~\cite{wong2016shisha,bovo2022cone,fidalgo2023magic}. Supporting multiple parallel views in a single user task also allows users to identify and localize target objects out of view more accurately in VR~\cite{teo2024evaluations,teo2023exploring}. However, these studies do not explore the ubiquitous yet fundamental task of objects' delivery or transfer in the shared view, and most of the proposed techniques only focus on recognizing the partner's point of interest. Moreover, there are limited discussion of existing techniques in multi-user environments. To the best of our knowledge, studies have rarely discussed the situation where the target object or collaborator is outside the user's FOV~\cite{fidalgo2023magic}. 
To address this gap, we propose a novel solution to enhance inter-user interaction by enabling users to manipulate and perceive objects within their collaborator's FOV, facilitated through a shared, controllable parallel view.

We introduce TeamPortal, which enables users to share and manipulate the objects of their collaborator's perspective through a view window. Users can manipulate or transfer objects within this window, facilitating seamless interaction. An initial user study demonstrated the effectiveness of TeamPortal for complex collaboration tasks, although some interaction scenarios are challenging. To address these limitations, we developed three variants, TeamPortal+, SnapTeamPortal+ and DropTeamPortal+, designed to mitigate motion sickness and latency by eliminating real-time view synchronization. Additionally, the latter two variants introduce distinct mechanisms for pausing the view window. We conducted a second user study to evaluate them.
The paper's contributions are: \textbf{(1)} prototypes supporting collaborative activities in VR with the aid of interactive parallel views, \textbf{TeamPortal and its three variants}; \textbf{(2)} the results of \textbf{two user studies} to evaluate the performance of these four technologies; and \textbf{(3)} \textbf{three design implications} from the study results to inform future VR collaboration system development.
\section{Related Work}
\paragraph{Object Transfer in Virtual Environment.}
Interaction design in virtual environments has increasingly focused on the seamless transfer, delivery, and sharing of objects between collaborators. The \textit{container} concept, proposed by Xia et al.~\cite{xia2018spacetime}, includes ``Parallel Objects'' and ``Avatar Objects'' to support parallel manipulation of objects and interaction between multiple users in virtual reality. Multiple users have the ability to control objects in this system without interference from others and share information about objects with each other by manipulating their avatars. 
Auda et al.'s pioneering work advances the area of object ownership transfer in virtual environments, in which passive haptic props facilitate two main strategies for managing virtual object ownership. Their investigation suggests that transferring strategies of virtual objects between remote users reduces the capability of communication and socialization but increases collaboration efficiency and fluency~\cite{auda2021m}. 
Similarly, Li et al.~\cite{li2019measuring}, and Fermoselle et al.~\cite{fermoselle2020let} conducted studies on sharing virtual objects in social VR. Specifically, Li et al. examined remote photo sharing, while Fermoselle et al. focused on the haptic simulation involved in the process of delivering documents. Although there has been some relevant research in this area, limited research specifically explores how to design sole visual user interfaces to improve multi-user experiences while reserving the efficiency and collaborative experience of transferring virtual objects. 

\paragraph{Shared Perspectives in Virtual Reality.}
Recent advances in AR and VR applications have demonstrated the significance of collaborative interactions through the appropriate use of visual cues and view sharing. Techniques that visualize gaze and improve point accuracy have been shown to reduce cognitive load and improve performance in AR and VR settings~\cite{li2019gaze,chen2021effect}.
Higuchi et al.,~\cite{higuch2016can} demonstrate the effectiveness of displaying collaborator's eye fixations for better coordination and attention focusing in remote tasks. Their research shows that clear visual cues in AR to enhance mutual understanding and efficiency in collaborative environments. 
In mixed reality (MR), adaptive avatars like \textit{Mini-Me} adjust to user behaviors to improve presence and interaction~\cite{piumsomboon2018mini}. Similarly, sharing gaze behaviors have been shown to enhance collaborative task performance by fostering a shared understanding, highlighting the importance of behavioral cues, represented visually by a \textit{Cone of Vision} for understanding attention dynamics in VR~\cite{jing2022the,bovo2022cone}.
Furthermore, the \textit{ShiSha} project leverages redirected avatars to facilitate face-to-face interactions within VR, enhancing the sense of presence and improving collaborative outcomes by building on the principles of direct gaze and visual cue effectiveness~\cite{Hoppe2021ShiSha}. Ablett et al.~\cite{ablett2023point} explored the \textit{Point \& Portal} technique, although they did not emphasize VR collaboration. Piumsomboon et al.~\cite{piumsomboon2019effects} examined the sharing of visual cues from the collaborator's FOV to highlight the attention to the effects of sharing awareness cues. Additionally, Piumsomboon et al.~\cite{piumsomboon2019shoulder} investigated sharing perspectives and videos with collaborators in the way of ``On the Shoulder of the Giant''.

These interconnected studies emphasize the transformative impact of visual cues and the sharing of view direction in advancing collaborative technologies in AR and VR. However, the existing works primarily focus on collaboration when users face the same direction, i.e., the user and the user's FOV are facing in the same direction or task. While the prior studies rarely involve situations where users are back-to-back or where users' fields of view do not overlap when collaborating, our solution of view sharing, characterized by manipulating features, serves as the first effort to address the issue. 

\paragraph{Parallel Views.}
One user's ability to control multiple avatars' virtual hands simultaneously is known as the \textit{Ninja hands} technique~\cite{schjerlund2021ninja}. This inspired the parallel views technique, which allows a user to view multiple views simultaneously. 
Teo et al. proposed \textit{NinjaHeads}~\cite{teo2023ninjaheads}, a parallel views system that generates four additional viewpoints in the four corners of the main view based on the user's gaze point. This allows the user to examine an object or a point of interest from different angles simultaneously, as if the user had more than one pair of eyes. Subsequently, Leo et al. developed three prototypes of parallel-view display and conducted user studies with simple and complex search tasks and distance estimation tasks. Their results show that parallel views can increase the efficiency of solving complex tasks and reduce the physical exertion of the user~\cite{teo2023exploring}. Also, Teo et al. ~\cite{teo2024evaluations} also investigated different numbers of additional views (2, 4, and 8) in the parallel views system. 
The \textit{OVRlap} technique~\cite{schjerlund2021ninja} also allows the user to perceive multiple locations simultaneously from a first-person perspective. 
However, they used a stacked transparent view, rather than overlaying multiple view windows on top of the main view, as implemented by Leo et al. Parallel views techniques have been validated in a number of prior single-user studies to improve task efficiency and user experience, especially for complex tasks and sizeable spatial contexts~\cite{teo2023exploring,teo2024evaluations,schjerlund2021ninja}. 
However, insufficient studies have discussed such techniques in a multiplayer collaborative environment, and we still lack knowledge about whether parallel views techniques can improve the efficiency and user experience of collaboration. 


\paragraph{Collaboration in Virtual Reality.}

Collaboration in VR has evolved as an important research direction, offering new possibilities for remote collaborative work. VR supports users to interact in a natural and intuitive way, provides enough shared visual workspace for users~\cite{fussell2000coordination}, and even provides a sense of users being together~\cite{schroeder2001collaborating}; these features allow VR to simulate collaborative work on site, which is more effective than sharing non-interactive 2D videos~\cite{macchi2024virtual}. 
The latest work has explored various dimensions for enhancing user collaboration in VR, including 3D scene reconstruction in VR to enable remote collaboration and exploring how different virtual cues (e.g., gaze~\cite{piumsomboon2017covar,higuch2016can,jing2022impact}, gestures~\cite{bai2020user,ullal2022multi,piumsomboon2018mini}, as well as pointers~\cite{teo2019investigating,wang2019head,wang2020view}, etc.). 
Despite these advances in VR collaboration, only a few studies focus on the impact of visualizing the views of other users in VR collaboration~\cite{bovo2022cone}, especially for interactable features of these views. 
Whereas shared views and grab behaviors have been shown in desktop collaboration environments to be an effective way to provide contextual information, which can increase users' awareness and understanding of their actions~\cite{fraser1999supporting,gaver1993one}. 
In contrast, our work primarily focuses on a shared view through a two-user collaborative VR setting. In particular, we design a view-sharing window that users can interact with. By designing variants of the shared view in our continual investigation, we investigate the generalized multi-user interaction techniques and evaluate their usability in complex tasks.

\section{TeamPortal: Design \& Implementation}
\label{System Design and Implementation}
Considering the research gap, we first designed the TeamPortal system, which provides a parallel view that can share and manipulate the collaborator's perspective. Figure~\ref{fig:teaser} gives a pictorial description of our system.


\subsection{Perspective Sharing and Manipulation Functions}

The two most important features of TeamPortal are (1) sharing the collaborator's FOV and (2) supporting direct interaction with the virtual objects in a parallel view showing the collaborator's FOV.
We constructed a semi-transparent view window in front of the user's perspective, which renders the contents of the collaborator's view in real time. This window design allows the user to focus on their own FOV and simultaneously observe the contents of the collaborator's FOV without obstruction. In addition, the user can use controller to move this view window within a certain range (within an FOV range of $0.5m$ to $2m$ between the user's eyes and the view window) to avoid occlusion. 
When the user wants to interact with a virtual object displayed in the view window (i.e., virtual objects in the view of the collaborator), simply press the ``A'' button (right hand) or the ``X'' button (left hand) on the controller, and the controller will ``shuttle'' into the view window for the user to perform the next operation. At this point, the FOV window will be rendered opaque so that the user can see the contents of the view window more clearly. The user uses distance grab to pick up objects both in their own FOV and in the view window, so that the user can more easily access objects in the far distance of their own and their collaborator's FOV. The user's controller will shoot a ray forward, when the user points the ray at the object they want to acquire, the object will turn blue (which means it is selected by the ray), at this point the user presses the controller's ``Grab'' button, the controller will grab the object, and the object will be placed when the button is released. 
When the user switches the ``A'' or ``X'' button on the controller while grabbing an object, the grabbed object can be switched between the view window and the user's own view. In this way, the user can acquire the object of interest from the collaborator's FOV or pass the object to the collaborator's FOV. This enables virtual objects to be transferred between two users through the view window.

To ensure seamless collaboration and avoid interaction conflicts, TeamPortal has a locking mechanism that controls object ownership. When a user interacts with an object (either in their own FOV or through the shared view window), the system automatically grants temporary ownership to that user. During this time, the object is ``locked'' preventing any other user from manipulating it until it is released. This locking mechanism ensures that only one user can operate an object at a time, eliminating potential conflicts caused by simultaneous user operations.

\subsection{System Implementation}

We developed the system using Unity (ver 2022.3.27f1) on a laptop equipped with a 12th Gen Intel Core i7-12650H processor, 32GB RAM, and an NVIDIA GeForce RTX 4070 GPU, running Windows 11. The participants used the Meta Quest 3 HMD, which supports a resolution of $2064 \times 2208$ pixels and a 120Hz refresh rate. User interactions were implemented using the Meta Interaction SDK, while multi-user networking was facilitated by Unity’s Netcode. Users could join the same virtual environment via the Meta Quest 3 HMD, navigating through teleportation~\cite{bozgeyikli2016point}.
\section{User Study 1: Shared View vs. Task Complexity}

We conducted a 3$\times$2 within-subjects design with
Technique (Baseline, ShaView, and TeamPortal) and Task Complexity (Simple and Complex; Section~\ref{sec:task_design}) as two independent variables. ShaView is a technique that allows view sharing without interactivity. In the Baseline condition,
a basic VR collaborative environment is only provided without a view window. The study involved participants completing two specific tasks under these different conditions. 

The first study is designed to evaluate the following hypotheses: \textbf{[H1]}: Users will perform collaborative tasks faster and more accurately with TeamPortal compared to Baseline and ShaView. \textbf{[H2]}: Perspective sharing (ShaView and TeamPortal) will enhance users' awareness of collaborator presence. 
\textbf{[H3]}: TeamPortal will reduce user movement by enabling object transfer through the view window. \textbf{[H4]}: TeamPortal will perform better on complex tasks due to its support for manipulating a parallel perspective.

\subsection{Task Environment and Design}
\label{sec:task_design}
\begin{figure}[htb]
  \centering
  \includegraphics[width=1\linewidth]{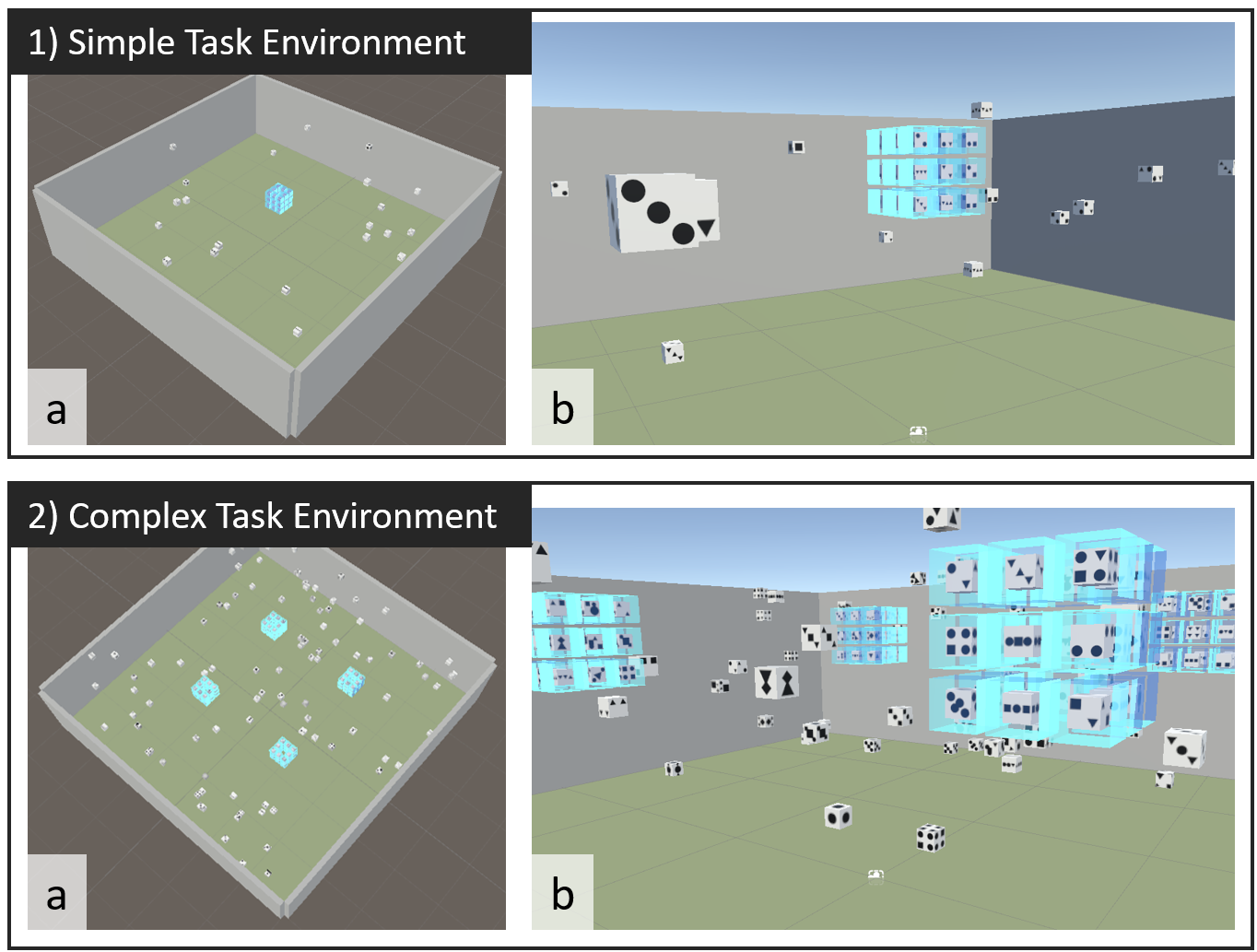}
  \caption{The overview of collaborative environments: Figures $1)a$ and $2)a$ present top views of simple and complex tasks, respectively; Figures $1)b$ and $2)b$ offer first-person perspectives within the collaborative environment for these tasks.}
    \vspace{-5mm}
  \label{fig:task_enviroment}
\end{figure}

We designed a 6m$\times$6m VR collaborative environment, which is a virtual room with a light green floor and gray exterior wall. The height of the wall is 1.8m, which represents the boundaries of the virtual world. Two collaborators operate inside the wall and can move freely around the green floor area via the teleport function, which is the task area of the collaborators.
We referred to the Cabinet task and Multi-Cabinet task~\cite{teo2023exploring} and designed two types of tasks, \textbf{simple tasks} and \textbf{complex tasks}. Also, we refer to the task that used the tangram shape to assess the performance of the parallel view~\cite{teo2024evaluations}, where the user navigates a 3D data cluster to identify a specific feature by carefully analyzing the attributes on each data surface. In a simple task (Figure~\ref{fig:task_enviroment}.~(1)), a target area composed of translucent light blue material cubes is generated in the center of the task area. These light blue cubes are arranged into a 3$\times$3 matrix formation, with the central column left empty. Each light blue cube contains a small hint at its center, which prompts the white cube tangram to match it. Accordingly, 24 suspended white cubes with a side length of 10cm will be generated at random locations in the 6m$\times$6m$\times$1.8m space of the collaboration area (the actual generation area is slightly smaller than this space size, this is to avoid the cubes being generated in the walls or stuck in the floor, and no cubes will be generated in the target area and the surrounding 0.5m). Each white cube features a unique tangram, with all six sides displaying the same design. This design enables users to view the tangram from any direction within the virtual room. The tangram of these 24 white cubes can be one-to-one matched with the target light blue cube. The collaborative environment for the complex task (Figure~\ref{fig:task_enviroment}.~(2)) is fundamentally similar to that for the simple task. However, the stacking area for the light blue target cubes has been expanded to include four distinct locations within the virtual room. 
In addition, the number of white cubes has increased to 96, with 24 cubes assigned to each of the four locations. Each white cube features a unique tangram, and each light blue target cube individually corresponds to one of the white cubes.

In summary, simple tasks represent fundamental interaction activities, where the user explores the environment and manipulates objects within a confined task area~\cite{teo2023exploring}. The concept involves: (1) the user moves to a task area, (2) searches for tangrams on 4 sides of the task area, and (3) locates the corresponding cube from a small set of scattered cubes (N = 24) in the virtual room, placing it in the appropriate position within the task area. In contrast, complex tasks simulate more advanced interaction processes, consisting of: (1) the user navigating across multiple task areas (N = 4), (2) searching for tangrams on the 16 sides of these task areas, and (3) identifying the corresponding cube from a larger set of scattered cubes (N = 96) in the virtual room and placing it in the appropriate location within the task area. In complex tasks, there will be more tangrams out of view in the target area due to occlusion.

In both simple and complex tasks, participants work in pairs to complete the task of matching cubes within the virtual room. They are required to match as many cubes as possible in 10 minutes, and they will be told that they can get more rewards if they complete the task better. Participants utilize the distance grab function~\cite{grab} (clicking and holding \textit{grab} button from a VR controller) to grip the white cube and move the white cube to the target area and find the corresponding position to drop the white cube (releasing the \textit{grab} button). Once a white cube is placed in the target area, the light blue material of the target cube will change to orange-red, signaling that a match has been made. However, participants will not receive any indication that the match is correct or incorrect. Participants can communicate during the study.

\subsection{Procedure, Measurement, and Participants}

Participants arrived at the study site in pairs at the scheduled time slot. First, they were asked to complete a demographic questionnaire, and then the researcher introduced the tasks and the functions of the TeamPortal prototype. Next, the researcher helped the participants wear the VR headset and allowed them to familiarize themselves with and practice using the TeamPortal prototype system in the test scene. Once participants were fully familiar with the operation of the system, they took a break from the headset for 1–2 minutes to prepare for the formal task. 

In the formal study, participants engaged in both simple and complex tasks under each of the three experimental conditions. The sequence of these conditions were counterbalanced using a Latin square design to mitigate order effects~\cite{keedwell2015latin}. Additionally, the order of simple and complex tasks within each condition were randomized to further control for any potential sequence-related biases. Each task was limited to a maximum duration of 10 minutes. 
Following completion of each task, participants had a 2-minute break. After completing both simple and complex tasks within a given condition, participants completed questionnaires to evaluate their experience of that specific condition. Upon completing all three conditions, the researcher interviewed the participants to discuss their preferences and their experiences during the study. The project was approved by the Hong Kong Polytechnic University ethics review board (Application No. HSEARS20240910004), and each participant would receive the equivalent of \$13 USD in drink store coupons at the end of the study.


To evaluate user task performance and experience across the three conditions, we employed both quantitative and qualitative analyses. For objective measurements, we recorded the number of cubes matched, the accuracy rate (\textit{Correctly matched cubes / Total placed cubes}), and the frequency of Teleport use. Additionally, we tracked participant movement, recording new positions if they moved more than 10 cm in the virtual environment. These positions were captured as 3D coordinates. 
For subjective assessments, we administered the NASA-TLX Scale (NASA-TLX) to measure subjective perceived workload~\cite{hart2006nasa}, the User Experience Questionnaire (UEQ) to evaluate user experience with the interactive product~\cite{schrepp2017design}, the System Usability Scale (SUS) to assess system usability~\cite{brooke1996sus}, and the Simulator Sickness Questionnaire (SSQ) to measure simulator and cybersickness~\cite{kennedy1993simulator}. We also used the Networked Minds Social Presence Measure to understand the participants' perceptions of partner interaction during the task~\cite{harms2004internal}. At the end of the study, participants were interviewed about their preferences and the reasons behind them.

We recruited 36 participants (18 pairs; 30 female, 6 male) from a local social media platform between the ages of 19 -- 38 years (M = 24.11, SD = 3.87). The majority of the participants (75\%) were in the age group of 22 -- 27 years. The familiarity of the participants with VR technology was measured on a 5-point scale, with 1 indicating no familiarity at all and 5 indicating a high level of familiarity. The mean VR familiarity score was 2.28 (SD = 0.85), but the majority of the participants had experienced VR devices before the study (83.3\%). 

\subsection{Results}
\label{Study1_Results}

We first assessed the internal consistency of our measurement instruments using Cronbach's Alpha to ensure reliable measurement of the constructs. To evaluate the effects of the three conditions, we conducted One-way Repeated Measures ANOVA (RM-ANOVA). Prior to performing the RM-ANOVA, we tested the normality of the residuals with the Shapiro-Wilk test and examined the assumption of sphericity using Mauchly's Test of Sphericity. If the sphericity assumption was violated, we applied the Greenhouse-Geisser correction to adjust the degrees of freedom. In instances where the normality assumption was not met, we employed non-parametric alternatives, i.e., the Friedman test, to appropriately analyze the data. Post hoc pairwise comparisons were conducted using paired t-tests when assumptions were satisfied or Wilcoxon signed-rank test multiple comparisons when assumptions were violated. These steps were consistently applied across all analyses to ensure the validity and reliability of our findings.

\begin{figure}[htb]
  \centering
  \includegraphics[width=1\linewidth]{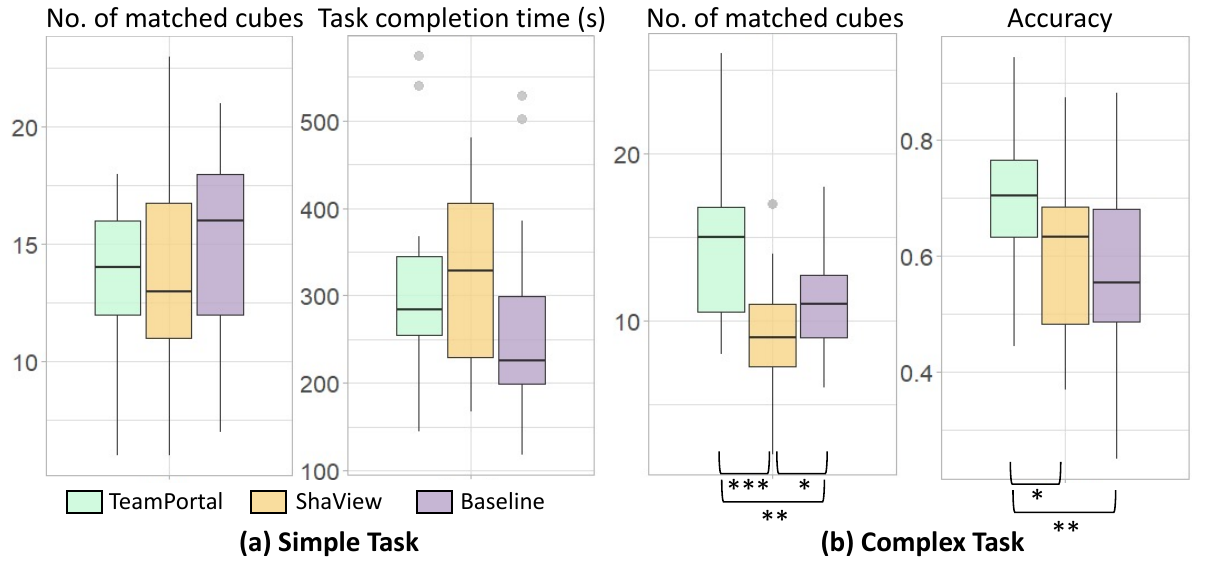}
  \caption{Task completion rate and accuracy of (1) simple task and (2) complex task ($P<.05$(*), $P<.01$(**), $P<.001$(***)).}
  \vspace{-5mm}
  \label{fig:Task_completion_rate}
\end{figure}

\paragraph{Task completion rate and accuracy}
\label{sec:task_completion_1}
During the \textbf{simple task}, all participants reported completion of the task in 10 minutes. Figure~\ref{fig:Task_completion_rate} (1) illustrates the number of cubes correctly matched by each group using the three different techniques, as well as the time taken to complete the task. There were no significant differences between techniques in either the number of cubes successfully matched ($F(1.5,25.5) = 1.35$, $p = .27$) or the time taken to complete the task ($\chi^2$(2) = 2.11, $p = .348$; Shapiro-Wilk test: $W = 0.87$, $p < .05$). Although the differences were not statistically significant, Baseline (M = 14.94, SD = 3.96) demonstrated slightly better performance in terms of the number of matched cubes compared to TeamPortal (M = 13.27, SD = 3.95) and ShaView (M = 14.05, SD = 4.99). Furthermore, Baseline (M = 264.78, SD = 114.08) was associated with a shorter average completion time than TeamPortal (M = 304.11, SD = 112.79) and ShaView (M = 322.78, SD = 108.18). This outcome may be attributed to the simplicity of the task, participants can quickly coordinate and identify cube patterns through direct communication without having to share a view window. In contrast, sharing the view window during a simple task might have introduced additional cognitive load, which could have led to increased time consumption (see Section~\ref{sec:dis_2} for more discussion).

In \textbf{complex tasks}, the three techniques had a significant impact on task completion rates ($F(2,34) = 17.89$, $p <.001$) and also accuracy ($F(2,34) = 5.17$, $p =.01$), Figure~\ref{fig:Task_completion_rate} (2) shows the box plot. Using TeamPortal resulted in significantly more successful cube matches compared to ShaView ($p<.001$) and Baseline ($p<.01$). In contrast, ShaView was the least effective, with significantly fewer successful matches than Baseline ($p<.05$). In addition, TeamPortal was the most accurate technology; it is significantly higher than ShaView ($p<.05$) and Baseline ($p<.01$). There are no significant differences between ShaView and Baseline ($p<.58$).

\begin{figure}[htb]
  \centering
  \includegraphics[width=1\linewidth]{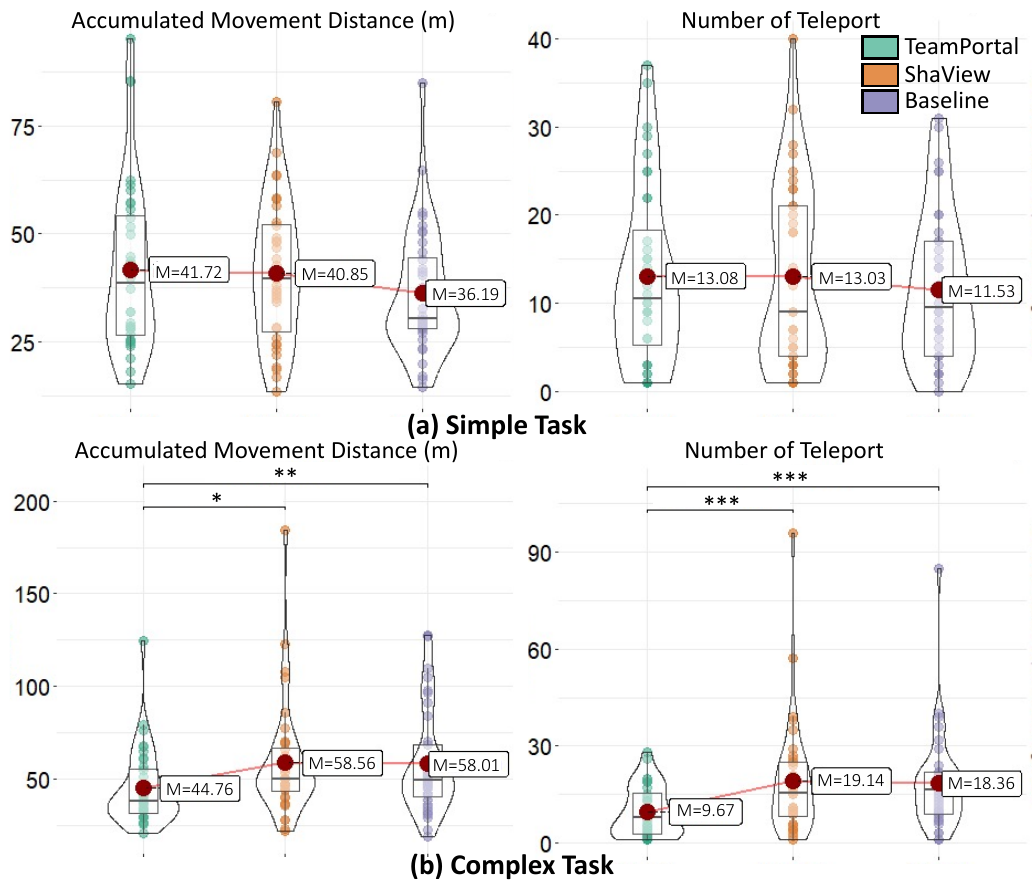}
  \caption{The accumulated movement distances and the number of times the participants used the Teleport function of (1) simple task and (2) complex task ($P<.05$(*), $P<.01$(**), $P<.001$(***)).}
  \vspace{-5mm}
  \label{fig:Behaviors_Study1}
\end{figure}

\paragraph{Comparison of user behaviors (Accumulated movement distance and Teleport counts)}
\label{sec:Userbehaviors1}
We calculated the movement of the accumulated distance ($\sum ^{n-1}_{i=1}={\sqrt {{({x}_{i+1}-{x}_{i})}^{2}+{({z}_{i+1}-{z}_{i})}^{2}}}$) and counted the number of movements of all participants using Teleport under the 3 conditions, the results are shown in Figure~\ref{fig:Behaviors_Study1}, there is no significant difference between the three cases in terms of the accumulated distance ($F(1.7,59.5) = 1.34$, $p=.27$) and the number of Teleports in the simple task ($F(1.7,59.5) = 0.42$, $p=.66$). However, in complex tasks, using the TeamPortal technique, participants will move significantly ($F(1.72,60.2) = 7.56$, $p<.01$) less than ShaView($p<.05$) and Baseline($p<.01$) and will use the Teleport function significantly fewer times ($F(2,70) = 13.05$, $p<.001$) than ShaView($p<.001$) and Baseline($p<.001$).

\paragraph{Subjective perceived workload (NASA-TLX)}

\begin{figure*}[htb]
  \centering
  \includegraphics[width=1\linewidth]{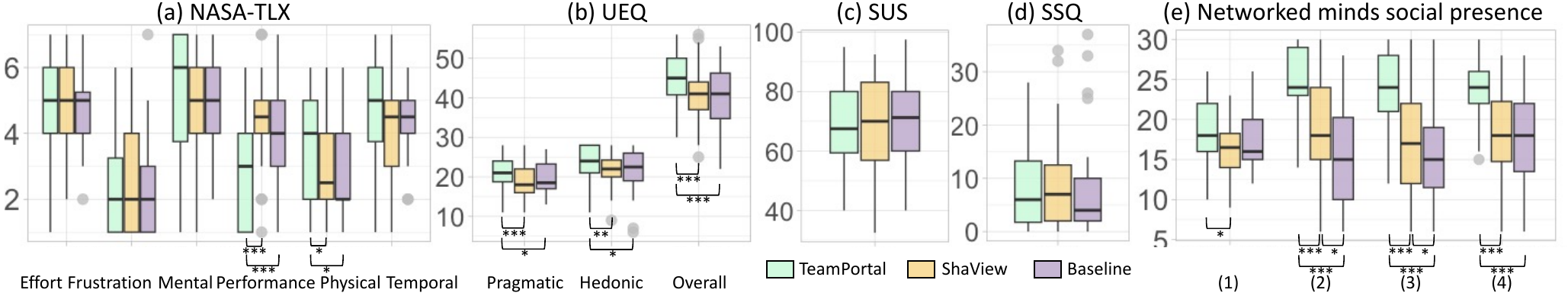}
  \caption{Box plots of subjective questionnaires for User Study 1. The four subscales of (e) are (1) Attentional allocation; (2) Co-presence; (3) Perceived behavioral interdependence and (4) Perceived message understanding ($P<.05$(*), $P<.01$(**), $P<.001$(***)).}
  \vspace{-5mm}
  \label{fig:Questionnaire1}
\end{figure*}

The results showed that there were no significant differences between the three Techs for the \textit{Mental demand} ($F(1.62,56.84) = 0.95$, $p=.38$), \textit{Temporal demand} ($F(2,70) = 0.70$, $p=.50$), \textit{Effort} ($F(2,70) = 0.99$, $p=.38$) and \textit{Frustration level} ($\chi^2$(2) = 1.07, $p = .587$; Shapiro-Wilk test: $W = 0.92$, $p < .05$) subscales, but there were significant differences for the \textit{Physical demand} (($F(2,70) = 5.3$, $p<.01$)) and \textit{Performance} (($F(2,70) = 15.13$, $p<.001$)) subscales (Figure~\ref{fig:Questionnaire1} (a)). 
TeamPortal increased \textit{Physical demand} significantly more than ShaView ($p<.05$) and Baseline ($p<.05$), which may be related to the fact that the techniques required more operations. In comparison, there is no significant difference between ShaView and Baseline ($p=.50$). Participants self-reported significantly better \textit{Performance} using TeamPortal than both ShaView ($p<.001$) and Baseline ($p<.001$), while there was no significant difference between ShaView and Baseline ($p=.21$), suggesting that using TeamPortal makes users feel significantly better about their performance.

\paragraph{User experience (UEQ)}

The RM-ANOVA results showed significant differences between the three Techs on the \textit{Pragmatic quality} ($F(2,70) = 9.41$, $p<.001$) and \textit{Hedonic quality} ($F(1.6,56) = 6.54$, $p<.01$) subscales and \textit{Overall} ($F(2,70) = 9.41$, $p<.001$) scores (Figure~\ref{fig:Questionnaire1} (b)). Post hoc analysis showed that TeamPortal had a significantly better \textit{Overall} user experience than ShaView ($p<.001$) and Baseline ($p<.001$). There was no significant difference between ShaView and Baseline ($p=.75$). \textit{Pragmatic quality} was significantly better for TeamPortal than for ShaView ($p<.001$) and Baseline ($p<.05$), and ShaView, although the mean value was lower than that of Baseline, but there is no significant difference ($p=.17$). TeamPortal is also significantly better than ShaView ($p<.01$) and Baseline ($p<.05$) in \textit{Hedonic quality}, and again, ShaView has a lower average \textit{Hedonic quality} than Baseline, but there is no significant difference ($p=.35$).

\paragraph{System Usability Scale (SUS)}

We evaluated the usability of three technologies using the SUS. The SUS is a standard tool for evaluating usability, providing a score between 0 and 100, where higher scores indicate better usability (Figure~\ref{fig:Questionnaire1} (c)). The average SUS scores were as follows: TeamPortal (M = 68.40, SD = 14.82), ShaView (M = 68.06, SD = 16.28) and Baseline (M = 69.72, SD = 14.83). Given that a score of 68 is generally considered the threshold for average usability~\cite{jordan1996usability}, all three Techs performed similarly. However, it is important to note that our participants were not very familiar with VR (see Section~\ref{Study1_Results}). Lack of experience could influence SUS scores~\cite{kortum2013usability}, potentially underestimating the usability of the systems due to the learning curve associated with new technology. Despite this, the SUS scores indicate that all three VR Techs provide an acceptable level of usability, suggesting that even for new users of VR, the systems are reasonably accessible.

\paragraph{Simulator Sickness Questionnaire (SSQ)}
\label{sec:SSQ1}
Simulator sickness was assessed using the total SSQ score after each session. The SSQ provides an overall measure of simulator sickness symptoms, with higher scores indicating greater severity (Figure~\ref{fig:Questionnaire1} (d)). 
Although Baseline (M = 7.75, SD = 8.99) has a slightly lower mean score than TeamPortal (M = 8.08, SD = 7.84) and ShaView (M = 8.61, SD = 8.72), the differences between the technologies are small and without significant differences ($\chi^2$(2) = 0.41, $p = .81$; Shapiro-Wilk test: $W = 0.84$, $p < .001$). This suggests that all three VR technologies cause similar levels of simulator sickness, and that their mean values range from 5-10, suggesting that users are producing minimal symptoms~\cite{kennedy2003configural}.

\paragraph{Networked minds social presence}

RM-ANOVA tests revealed significant differences for all measures (Co-presence: $F(2,70) = 35.64$, $p<.001$; Attentional allocation: $F(2,70) = 3.64$, $p<.05$; Perc. message understanding: $F(2,70) = 41.52$, $p<.001$; Perc. behavioral interdependence: $F(2,70) = 40.44$, $p<.001$; see Figure~\ref{fig:Questionnaire1} (e)). Manual post hoc pairwise comparisons by each dimension indicated that TeamPortal significantly enhanced the \textit{Co-presence} of both ShaView ($p<.001$) and Baseline ($p<.001$). A significant difference was also observed between ShaView and Baseline ($p<.01$), suggesting that both TeamPortal and ShaView significantly improve \textit{Co-presence}, and that TeamPortal is significantly better than ShaView. For the \textit{Attentional allocation} dimension, we found that only TeamPortal was significantly better than ShaView ($p<.05$). 
This suggests that TeamPortal significantly enhances attention between collaborators compared to ShaView and that collaborators have more difficulty reducing attentional connectivity due to the influence of the environment or task. At the level of \textit{Perceived message understanding}, TeamPortal is significantly better than ShaView ($p<.001$) and Baseline ($p<.001$), but there is no significant difference between ShaView and Baseline ($p=.09$). This suggests that TeamPortal improves the ability to \textit{Perceived message understanding} between collaborators. For the \textit{Perceived behavioral interdependence} scale, TeamPortal is significantly more dependent on the collaborator compared to both ShaView ($p<.001$) and Baseline ($p<.001$). ShaView also depends more significantly on the collaborator than Baseline ($p<.05$). This suggests that both TeamPortal and ShaView significantly increase behavioral dependence between partners; however, TeamPortal makes the dependence significantly stronger.

\paragraph{Reliability Assessment}
\label{sec:Reliability1}
We assessed the internal consistency of each standardized questionnaire using Cronbach’s alpha.  NASA-TLX exhibited ($\alpha=0.74$), UEQ ($\alpha=0.78$), SUS ($\alpha=0.53$), SSQ ($\alpha=0.92$) and Networked Minds Social Presence Scale ($\alpha=0.87$). Following common guidelines ($\alpha \geq 0.70$ as acceptable~\cite{nunnally1994psychometric}), these results indicate that the questionnaires demonstrated acceptable to high reliability except SUS, while the SUS score fell below the recommended threshold in our sample. Therefore, we recommend interpreting the SUS-related findings with caution.

\paragraph{Preferences and subjective comments}
\label{subjective_comments1}
The vast majority of the participants (N = 30, 83.33\%) rated TeamPortal as the best condition, whereas more than half of the participants (N = 22, 61.11\%) thought ShaView was the worst condition. In addition to that, some participants (N = 6, 16.67\%) considered that Baseline was the best condition. However, none of the participants thought that ShaView was the best condition. We scored the 3 conditions according to the preferences of the participants (1: Worst - 3: Best), the results of Friedman test ($\chi^2$(2) = 39.5, $p < .001$; Shapiro-Wilk test: $W = 0.63$, $p < .001$) indicated that TeamPortal scored significantly higher than ShaView ($p<.001$) and Baseline ($p<.001$), however, there was no significant difference between ShaView and Baseline ($p=.09$).

We asked the participants their reasons for ranking, and all the participants who ranked TeamPortal first said that the technology was much more convenient than ShaView and Baseline. Some participants (N = 5) mentioned that they did not need this technology much for simple tasks. Almost half (N = 16) of the participants stated that they thought TeamPortal would be more helpful if they could pause the view window while selecting an object. As we observed the participants doing the task, they consistently communicated over and over again, such as, ``\textit{turn your head a little bit to the left...}'', ``\textit{...too much, a little bit more to the right.}'', ``\textit{keep your head still! I'm going to take a cube.}'' or ``\textit{Can you move your head a little slower?}''. Some participants (n = 2) noted that it was not very convenient to use the view window to place the cube into the target area in their partner's FOV (we observed only 2 participants using the view window in this way), stating that because the view window updates the partner's FOV in real time, they could only place the object into the correct target cube in the partner's FOV from the view window if the partner kept their head perfectly still. Participants who ranked ShaView last overwhelmingly indicated that this view window was not helpful or even frustrating to them in completing the task. Two participants said they saw the cube they needed in the view window, but felt anxious and rushed because they could not access it directly. One participant said: ``\textit{I prefer not to be allowed to see the cube in the window, it would be difficult for me to describe the pattern to my partner and if I move around to look for it myself, I would be disoriented.}'' Some participants (N = 4) indicated that having only the view window and not being able to interact would make them feel more motion sickness. The reason some participants (N = 13) indicated that ShaView was inferior to Baseline was that instead of providing them with benefits, ShaView would increase their cognitive load.
\section{User Study 2: Variants of TeamPortal}

\begin{figure*}[htb]
  \centering
  \includegraphics[width=1\linewidth]{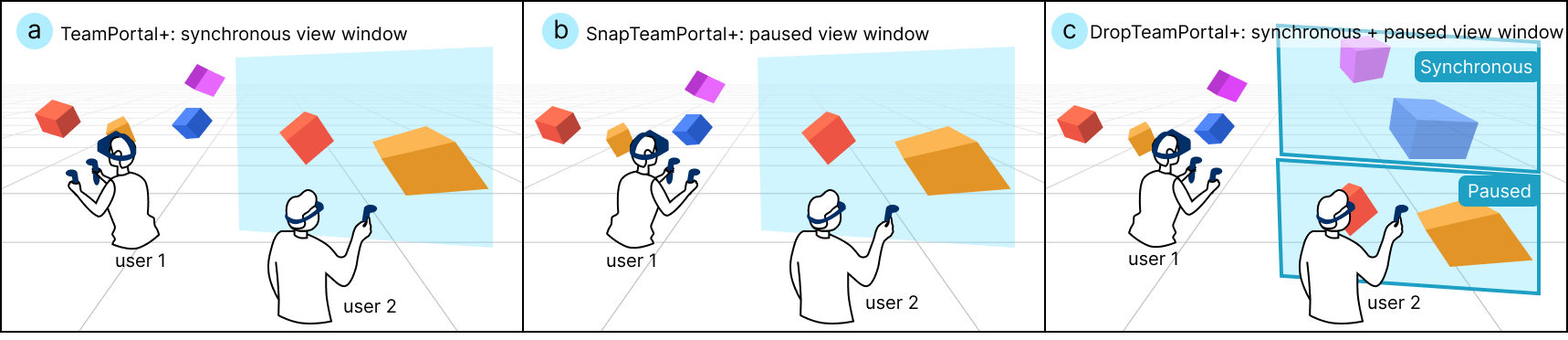}
  \caption{Illustration of the TeamPortal variants; (1) TeamPortal+; (2) SnapTeamPortal+; and (3) DropTeamPortal+.}
   \vspace{-5mm}
  \label{fig:Variants_examples}
\end{figure*}

The first user study generates valuable cues for designing collaborative views in two levels of task complexity. In particular, we highlight our observations, as follows: We know that \textbf{(1)} In complex tasks, the use of TeamPortal leads to significant improvements in task completion rates and accuracy. However, in simple tasks, no objective advantage is observed when using TeamPortal (see Section~\ref{sec:task_completion_1}).
\textbf{(2)} Real-time synchronization of view windows may increase motion sickness, as some participants (N = 4) reported feeling dizzy during the experiment, despite the SSQ scores being within acceptable ranges (see Section~\ref{sec:SSQ1} and~\ref{subjective_comments1}). \textbf{(3)} The utilization of TeamPortal increases the communication costs for users due to the additional effort required to moderate their FOV (see Section~\ref{subjective_comments1}). \textbf{(4)} In some specific usage cases, synchronization of the direct FOV for TeamPortal may be difficult for the user to operate (see Section~\ref{subjective_comments1}). 
Accordingly, our first study demonstrated the necessity of shared views for user collaboration, especially when the task complexity grows. From the above analysis, we have made the following improvements to improve the usability and comfort of TeamPortal: \textbf{(A)} Perspective synchronization is triggered only when head movement exceeds 10 cm or head rotation surpasses 5\textdegree (a preliminary study with 4 participants determines the thresholds), with interpolation applied to ensure smooth transitions. This approach reduces unnecessary updates, thus optimizing network performance and alleviating motion sickness. Additionally, it lessens the need for precise head positioning, reducing user workload. \textbf{(B)} The option of pausing the view window while interacting with TeamPortal is implemented. This is accomplished by placing a virtual camera aligned with the partner's view.

It is worth highlighting that TeamPortal has demonstrated significant advantages in complex task environments. Thus, User Study 2 follows the same setup of \textbf{complex tasks} as User Study 1. We conduct an in-depth study to evaluate the variants of TeamPortal, namely TeamPortal+, SnapTeamPortal+, and DropTeamPortal+. 
We named the upgraded perspective synchronization version \textit{TeamPortal+}, and designed two types of pause-enable interfaces (Figure~\ref{fig:Variants_examples} for details). The first interface allows users to enable pauses of view sharing directly within the TeamPortal+ view window (\textit{SnapTeamPortal+}). 
When pausing, the second interface initiates the opening of a new window immediately below the original TeamPortal+ window. The newly created window then shows the paused status. 
When user interaction is no longer required, the newly opened pause window will disappear (\textit{DropTeamPortal+}). 

In this study, our objective is to evaluate the following hypotheses: \textbf{[H5]}: SnapTeamPortal+ and DropTeamPortal+ will increase task completion rates by reducing task-independent communication through pausing view sharing during interactions within the view window. \textbf{[H6]}: SnapTeamPortal+ will reduce users' awareness of collaborator presence due to the pause in view sharing while interacting with the view window. \textbf{[H7]}: DropTeamPortal+ will be the most preferred by users, as it supports both perspective synchronization and pausing view sharing during interactions. \textbf{[H8]}: The frequency of using the view window to transfer objects will positively impact the task completion rate.

\subsection{Participants, Study Design and Measurement}
Another 36 participants (18 pairs; 29 female, 7 male) were recruited from a social media platform, and the participants were aged 20 to 35 years (M = 24.22, SD = 2.80). Their mean value of VR familiarity was 2.19 (SD = 0.91) on a 5-point scale. 63.89\% of the participants had used the VR device before.
We used the same study procedure, design, and measurement as stated in the previous study, except that participants were only required to complete the complex task under each condition, and each task was extended to 15 minutes. Thus, the study becomes a 1$\times$3 within-subjects design. In other words, each participant pair was required to perform the task three times using different techniques.

\subsection{Result}
Our User Study 2 used the same process for data analysis as User Study 1, with details given at the beginning of Section~\ref{Study1_Results}.

\begin{figure}[htb]
  \centering
  \includegraphics[width=1\linewidth]{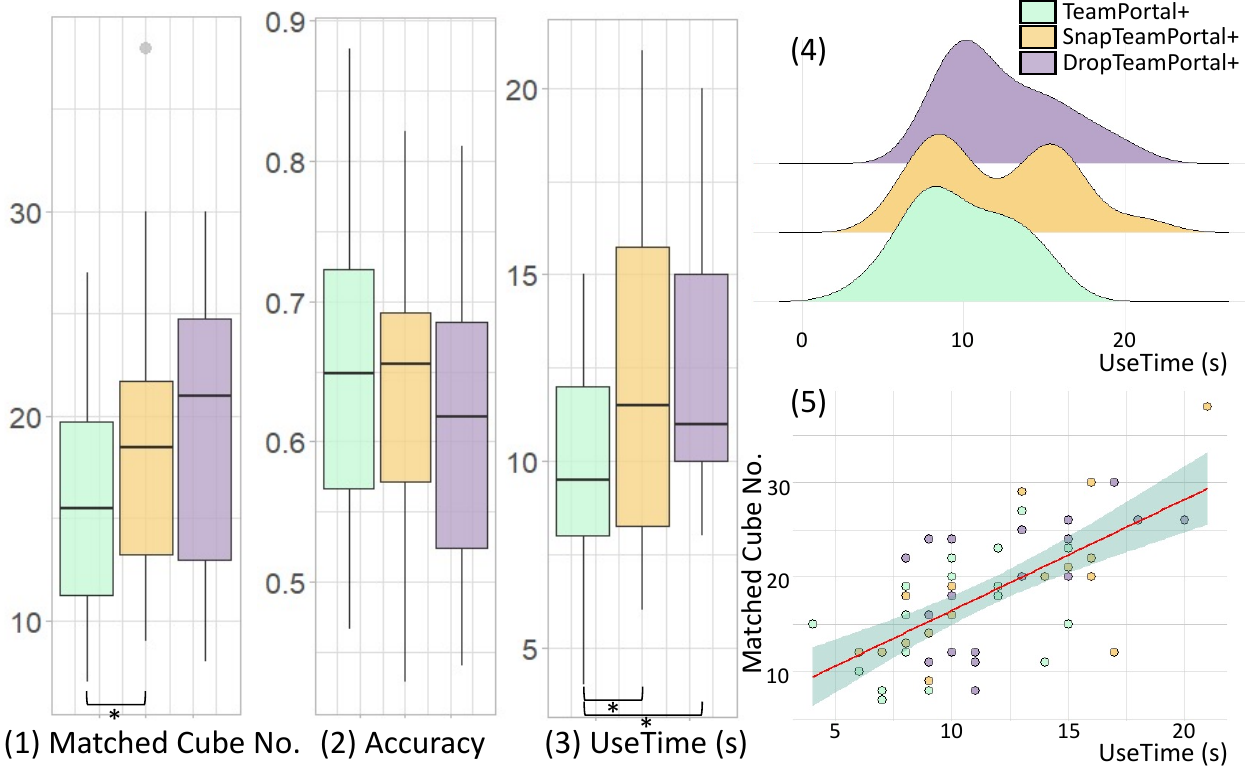}
  \caption{Number of successfully matched cubes, accuracy and number of view window interactions (UseTime) of variants ((1) to (3)); (4) density distribution of UseTime; and (5) UseTime vs. matched cubes with linear regression ($P<.05$(*), $P<.01$(**), $P<.001$(***)).}
  \vspace{-3mm}
  \label{fig:Behaviors_Study2}
\end{figure}

\paragraph{Task completion rate and accuracy}
As shown in Figure~\ref{fig:Behaviors_Study2} (1) and (2), the three techniques can significantly impact task completion rates ($F(2,34)=3.64$, $p<.05$). Both SnapTeamPortal+ (M=19.72, SD=6.59) and DropTeamPortal+ (M=19.11, SD=7.44) increase the number of correctly matched cubes on average compared to TeamProtal+ (M=15.83, SD=5.92). Post hoc analyzes reveal that SnapTeamPortal+ has a significant effect on completion rates ($p<.05$). However, none of the three techniques significantly affect accuracy ($F(2,34)=0.48$, $p=.62$).

\paragraph{Comparison of user behaviors}
\label{sec:Userbehaviors2}
The total number of cube passes made by each group of participants within the view window was evaluated across the three different techniques. The findings show that both the SnapTeamPortal+ ($p<.05$) and DropTeamPortal+ ($p<.05$) techniques significantly enhanced participants' propensity for use ($F(2,34)=3.64$, $p<.05$), resulting in a higher frequency of cube passes the view window (Figure~\ref{fig:Behaviors_Study2} (3)). Figure~\ref{fig:Behaviors_Study2} (5) illustrates a scatter plot showing the relationship between the number of uses (UseTime) and the number of successfully matched cubes for the three techniques. A linear regression analysis was performed on the data, revealing a positive linear relationship between the frequency of use and the number of successful matches. This trend suggests that as participants engage more frequently with these techniques, their success rate in matching cubes tends to increase. Furthermore, a bimodal distribution was observed in the frequency of SnapTeamPortal+ usage (Figure~\ref{fig:Behaviors_Study2} (4)), indicating a divergence in user preference for this technology. This finding is supported by subjective feedback from participants, a more detailed discussion of this finding, along with relevant user comments, will be presented in Section~\ref{sec:subjective2}. Consistent with User Study 1, we calculated both the total distance moved by participants and the number of times the Teleport function was used. No significant differences exist in either metric. 

\paragraph{Subjective perceived workload (NASA-TLX)}

\begin{figure*}[htb]
  \centering
  \includegraphics[width=1\linewidth]{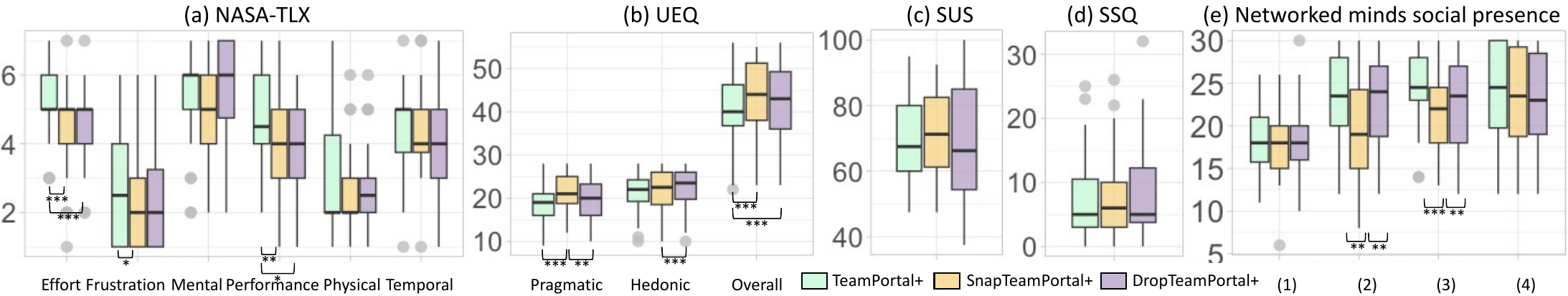}
  \caption{Box plots of subjective questionnaires for User Study 2. The four subscales of (e) are (1) Attentional allocation; (2) Co-presence; (3) Perceived behavioral interdependence and (4) Perceived message understanding ($P<.05$(*), $P<.01$(**), $P<.001$(***)).}
  \vspace{-5mm}
  \label{fig:Questionnaire2}
\end{figure*}

A series of RM-ANOVA showed that there were no significant differences between the three conditions for mental demand ($F(2,70)=2.65$, $p=.08$), physical demand ($F(2,70)=2.31$, $p=.11$), and temporal demand ($F(2,70) = 0.47$, $p=.63$). However, a significant effect of technology was found for performance ($F(2,70)=6.11$, $p<.01$), effort ($F(2,70)=15.38$, $p<.001$), and frustration ($\chi^2$(2)=7.39, $p<.05$; Shapiro-Wilk test: $W=0.93$, $p<.05$; (Figure~\ref{fig:Questionnaire2} (a)). Post hoc pairwise comparisons showed that SnapTeamPortal+ ($p<.01$) and DropTeamPortal+ ($p<.05$) were significantly better than TeamPortal+ in terms of performance, but there was no significant difference between SnapTeamPortal+ and DropTeamPortal+ ($p=.53$). Similarly, using SnapTeamPortal+ ($p<.001$) and DropTeamPortal+ ($p<.001$) requires significantly less effort than TeamPortal+. However, the difference between SnapTeamPortal+ and DropTeamPortal+ is not significant ($p=.52$). Besides that, there was a significant reduction in frustration with SnapTeamPortal+ compared to TeamPortal+ (($p<.05$)), while there was no significant difference between the other conditions.

\paragraph{User experience (UEQ)}

In general, TeamPortal+ had significantly lower UEQ scores than both SnapTeamPortal+ ($F(2,70)=15.26$, $p<.001$; $p<.001$) and DropTeamPortal+ ($p<.001$), with no significant difference observed between DropTeamPortal+ and DropTeamPortal+ ($p=.13$; Figure~\ref{fig:Questionnaire2} (c)). Further analysis revealed that SnapTeamPortal+ scored significantly higher in Pragmatic Quality ($F(2,70)=15.26$, $p<.001$) compared to both TeamPortal+ ($p<.001$) and DropTeamPortal+ ($p<.01$). Conversely, DropTeamPortal+ showed significantly higher Hedonic Quality compared to TeamPortal+ ($F(1.64,57.4)=6.28$, $p<.01$; $p<.001$). These findings suggest that although there was no significant difference in overall UEQ scores between SnapTeamPortal+ and DropTeamPortal+, SnapTeamPortal+ outperformed in terms of utility (Pragmatic Quality), while DropTeamPortal+ excelled in terms of entertainment (Hedonic Quality).

\paragraph{System Usability Scale (SUS)}

The results of our analysis show that the mean SUS scores for the three technologies are greater than 68, indicating that all three of our systems meet the average usability threshold (Figure~\ref{fig:Questionnaire2} (c)). The mean score for SnapTeamPortal+ (M = 70.90, SD = 13.42) is slightly higher than that of TeamPortal+ (M = 68.89, SD = 12.64) and DropTeamPortal+ (M = 68.61, SD = 16.84), but there are no significant differences ($F(1.66,58.1)=0.54$, $p=.58$). In addition, we found that all three variants had slightly higher mean SUS scores than TeamPortal (M = 48.40), which may potentially indicate a benefit to our optimized approach of perspective synchronization.

\paragraph{Simulator Sickness Questionnaire (SSQ)}

Similarly to User Study 1, we calculated total scores on the SSQ scale (Figure~\ref{fig:Questionnaire2} (d)). The mean scores for SnapTeamPortal+ (M = 7.47, SD = 6.59) were slightly lower than those for TeamPortal+ (M = 7.92, SD = 6.59) and DropTeamPortal+ (M = 8.27, SD = 7.39), but there were no significant differences between conditions ($F(2,70)=0.54$, $p=.59$). The mean value for the three conditions ranged from 5 to 10, suggesting that users produce minimal symptoms.

\paragraph{Networked minds social presence}
\label{sec:presence2}
The RM-ANOVA revealed significant differences among the three techniques on the Co-presence ($F(1.66,58.1)=11.39$, $p<.001$) and Perceived Behavioral Interdependence ($F(2,70)=10.74$, $p<.001$) subscales. However, no significant differences were observed in the Attentional Allocation ($F(2,70)=0.44$, $p=.64$) and Perceived Message Understanding ($F(1.54,53.9)=0.21$, $p=.81$) subscales (Figure~\ref{fig:Questionnaire2} (e)). Post hoc analyses indicated that SnapTeamPortal+ had significantly lower scores in Co-presence compared to both TeamPortal+ ($p<.01$) and DropTeamPortal+ ($p<.01$). Additionally, TeamPortal+ showed significantly higher Perceived Behavioral Interdependence scores than both SnapTeamPortal+ ($p<.001$) and DropTeamPortal+ ($p<.01$). These results suggest that SnapTeamPortal+ may reduce the sense of co-presence among collaborators, while DropTeamPortal+ maintains a similar level of co-presence as TeamPortal+. Furthermore, the stronger behavioral interdependence observed in TeamPortal+ may be attributed to the need for frequent communication between collaborators to continuously adjust the view window (Section~\ref{subjective_comments1}).

\paragraph{Reliability Assessment}

Similarly to Study 1 (Section~\ref{sec:Reliability1}), NASA-TLX ($\alpha=0.74$), UEQ ($\alpha=0.88$), SUS ($\alpha=0.59$), SSQ ($\alpha=0.89$) and the Networked Minds Social Presence Scale ($\alpha=0.78$) results indicate that these questionnaires demonstrated acceptable to high reliability, although we may need to be cautious about SUS-related findings.

\paragraph{Preferences and subjective comments}
\label{sec:subjective2}

More than half of the participants rated TeamPortal+ as the least favorable technology (N=23, 63.89\%). In contrast, approximately half rated SnapTeamPortal+ (N=18, 50\%) and DropTeamPortal+ (N=15, 41.67\%) as providing the best experience, with only two participants (N=2) selecting TeamPortal+ as the best technology. The results of Friedman test and post hoc analyses revealed that significantly more participants preferred SnapTeamPortal+ ($\chi^2$(2)=17.64, $p<.001$; Shapiro-Wilk test: $W=0.21$, $p<.001$; $p<.001$) and DropTeamPortal+ ($p<.001$) over TeamPortal+. However, no significant difference was found between SnapTeamPortal+ and DropTeamPortal+ ($p=.95$).

User feedback revealed an interesting polarization regarding SnapTeamPortal+. Some participants found SnapTeamPortal+ the most convenient technology, appreciating that they could interact with the view window to pick up the target cube simply. They also found the single-window design intuitive, noting that it did not impose additional cognitive load. However, other participants ranked SnapTeamPortal+ as the least favorable, citing a loss of co-presence during interactions. One user noted, \textit{``I can't tell where he (partner) is looking when I use it, so I can't confirm whether we are looking the cube simultaneously. It felt like a solo game for those few seconds.''} DropTeamPortal+ received generally positive feedback, with participants appreciating the retention of a shared perspective and the ability to pause during operations for easier interaction. However, some participants expressed concerns about the increased cognitive load due to the two-window design. One participant said \textit{``...My own view, her (partner) view in the window, and the paused view... I don't think I can manage looking at that many screens.''} 
Also, none of the participants in this user study reported feeling dizzy or having difficulty controlling their head movements. Furthermore, no participants indicated that smooth transitions caused them to lose track of their partner’s view. As one participant commented, \textit{``I didn't notice any trigger threshold for the view window, I thought it was my partner's head spinning slow and steady.''}
\section{Discussion}
\paragraph{User Collaboration: Manipulation or Observation?}
Our findings suggest enabling manipulation is more advantageous than sole observation in user collaboration. In our first user study, we compared three conditions: Baseline, ShaView, and TeamPortal. The results demonstrated that ShaView performed the worst in complex tasks, both in terms of task completion rate and user preference, contradicting \textbf{[H2]}. Interestingly, this aligns with findings from Teo et al., whose study on Parallel Views in a single-user task revealed that the absence of Parallel Views led to better user performance~\cite{teo2023exploring}. Conversely, TeamPortal emerged as the optimal condition, with higher task completion rates and user preference, supporting \textbf{[H1]}. The key distinction between ShaView and TeamPortal lies in the ability to interact with the view window. This finding suggests that direct manipulation of the view window is essential for effective collaboration. Merely sharing an observable view window without interactive control may reduce collaboration efficiency rather than enhance it. A possible explanation for this phenomenon is that the ability to manipulate the view window not only enables users to access what they need more directly but also increases their sense of engagement in the task~\cite{christopoulos2018increasing}.

\paragraph{TeamPortal in Simple vs. Complex Tasks}
\label{sec:dis_2}
Our study found that TeamPortal excelled only in complex tasks, with most participants reporting minimal or no use of the view window for object manipulation and transfer during simpler tasks, supporting \textbf{[H4]}. In simpler tasks, participants indicated that they could effectively divide responsibilities. For example, one participant might handle two sides of the target area (out of four), or one would manage cube manipulation while the other monitored all sides and placed the cubes. The relatively low complexity of tasks involving 24 cubes made task division straightforward, rendering TeamPortal unnecessary. Moreover, participants found that using TeamPortal introduced additional steps that slowed progress. For instance, to use the view window, they first had to communicate with their partner to identify the cube needed, then enter the window to retrieve it, while their partner had to maintain the correct orientation. These interactions were perceived as redundant and, in some cases, even impeded task efficiency in simple scenarios.

In contrast, TeamPortal demonstrated highly advantageous performance in complex environments. As depicted in Section~\ref{sec:Userbehaviors1}, users using TeamPortal significantly reduced their use of Teleport and overall movement, as they could manipulate objects directly through the View Window, which supports \textbf{[H3]}. Previous research has consistently shown that instantaneous movement methods in VR, such as Teleportation, can impair spatial perception and lead to disorientation due to the lack of continuous motion observation \cite{rahimi2018scene,bowman1998methodology,bowman1999maintaining}. In complex tasks, participants frequently encountered situations in which they identified a target but became disoriented after teleporting to retrieve it. TeamPortal mitigates this issue by allowing users to interact with objects without changing their physical location, reducing the likelihood of losing orientation. In such scenarios, the additional communication required to use TeamPortal is outweighed by the significant benefits of reduced movement and improved task efficiency.

\paragraph{Trade-offs Between TeamPortal Variants}

The second user study compares three variants of TeamPortal: TeamPortal+, SnapTeamPortal+, and DropTeamPortal+. We observe trade-offs between efficiency and collaboration experience. The results show that both SnapTeamPortal+ and DropTeamPortal+ improved task completion rates and significantly increased users' willingness to use them, consistent with \textbf{[H5]}. Additionally, SnapTeamPortal+ was found to reduce the sense of co-presence among collaborators, in line with \textbf{[H6]}. Furthermore, the usage frequency of view windows is positively correlated with the number of successfully matched cubes, which is consistent with \textbf{[H8]}. However, contrary to \textbf{[H7]}, DropTeamPortal+ was not the most preferred technology among users. 
These outcomes appear to be linked to trade-offs between SnapTeamPortal+ and DropTeamPortal+. As discussed in Section~\ref{sec:Userbehaviors2} and~\ref{sec:subjective2}, SnapTeamPortal+ displayed a bimodal distribution of user preferences, indicating polarized opinions. Users who prioritized task success favored SnapTeamPortal+ due to its simplicity (with only one window) and the ability to pause perspective sharing, which minimized task-irrelevant communication. For these users, DropTeamPortal+ was perceived as unnecessarily complex. However, users who were more focused on collaboration and real-time communication found SnapTeamPortal+ has limitations, particularly the temporary loss of their partner’s view during interactions. The ``Pause'' function may also lose the sense of co-presence (Section~\ref{sec:presence2}), which is unacceptable to this group of users. This part of the users felt that SnapTeamPortal+ was less effective than even TeamPortal+, which does not pause perspective sharing. Although DropTeamPortal+ had the highest average usage, suggesting broad acceptance, it also has challenges, such as increased cognitive load due to the presence of multiple windows~\cite{kueker2024learning}.

\paragraph{Design Implications}

Our studies generate design clues with three implications for shared and manipulating parallel perspective interfaces in VR collaboration. 
\textbf{DI1:} Offer Multiple View Window Modes. Users should be able to switch between modes that prioritize task performance (e.g., reduced communication, directly pause the view window [SnapTeamPortal+]) or collaboration (e.g., ongoing perspective sharing [TeamPortal+ or DropTeamPortal+]) depending on task complexity and task purpose. 
\textbf{DI2:} Adapt to Task Complexity. For simple tasks, parallel perspectives may be unnecessary, while in complex tasks, TeamPortal and its variants offer significant advantages. 
\textbf{DI3:} Consider Collaborative Virtual Environments (CVE) Size. TeamPortal has the advantage of reducing the need for teleportation, particularly in larger virtual environments. As the size of the CVE increases, TeamPortal and its variants should be prioritized to minimize user movement and enhance spatial awareness. 

\paragraph{Limitations and Future Work}

In both studies, participants from local social media platforms had minimal familiarity with VR systems, with some having no prior VR experience. This may have influenced the evaluation of our prototypes. For example, users with more VR experience might prefer DropTeamPortal+ due to its additional functionality, while SnapTeamPortal+ is simpler to operate. Further research is needed to examine how VR experience affects these preferences. Additionally, it may also affect the SUS score, as we pointed out in Section~\ref{Study1_Results}. However, our study addresses a limitation in Teo et al.~\cite{teo2024evaluations}, which focused only on VR-experienced users. Our findings show that even participants with limited VR experience were able to quickly learn and effectively use TeamPortal and its pauseable variants, SnapTeamPortal+ and DropTeamPortal+, suggesting that the system is accessible to a broad range of users. 
Another notable limitation of our study is that we did not compare TeamPortal and its variants with systems that support visual cue sharing, such as annotation or pointing~\cite{chen2021effect,higuch2016can,piumsomboon2018mini,bovo2022cone}. This decision was based on two considerations. First, visual cue sharing is most effective when users’ fields of view overlap; in scenarios where users are positioned back-to-back, shared annotations may be less useful. However, this assumption requires further investigation. Future studies could explore adding shared annotations or pointing to the view window, enabling users to see each other’s views and gestures. Second, the study duration was already extensive, with each participant pair spending nearly two hours in total. Adding more comparisons, such as shared visual cues, would have further lengthened the experiment. We prioritized evaluating the interaction aspects most relevant to our design, particularly view window manipulation and observation. Nonetheless, future work should investigate the comparative effects of shared visual cues and shared perspectives. For the system design, the view window of our TeamPortal, as well as its variants, uses flat, two-dimensional graphic rendering, which may lose depth information. Nonetheless, real-time stereoscopic view window rendering may exacerbate bandwidth demands, hence heightening latency and perhaps resulting in an unpleasant user experience. Future studies may need further assessment of these design issues.

For further studies, a potential research direction involves extending TeamPortal’s manipulable view-sharing technology to support multiplayer scenarios. Such an extension raises questions about system design and user interaction. For example, how should a multiplayer view window interface be organized and displayed? Should multiple parallel views be presented simultaneously~\cite{teo2023exploring,teo2024evaluations}, or should users switch among individual view windows (similar to the OVRlap technique~\cite{schjerlund2021ninja})? Additionally, the number of collaborators that interact with TeamPortal at once can influence its overall effectiveness, as an excessive number of view windows could increase cognitive load and potentially confuse users. Investigating these issues will provide valuable insights for extending and promoting TeamPortal for larger, more complex multi-user environments.
\section{Conclusion}
In this paper, we designed and implemented TeamPortal, a technology for sharing and manipulating parallel perspectives in collaborative tasks, and evaluated it in the first user study on both simple and complex tasks. Based on these results, we developed three variants (TeamPortal+, SnapTeamPortal+, and DropTeamPortal+), with the latter two allowing pause perspective sharing during manipulation. A second user study assessed the usability of these variants and highlighted trade-offs for different users. Our findings demonstrate the effectiveness of TeamPortal and its variants in complex tasks and provide recommendations for designing parallel perspective visualization techniques for VR collaboration.


\acknowledgments{%
This research was supported by the Hong Kong Polytechnic University's Start-up Fund for New Recruits (Project ID: P0046056) and the Science Research Project of Hebei Education Department, China (Project ID: QN2025723). Xian Wang was supported by a grant from the PolyU Research Committee (code RMHD). 
}

\newpage

\bibliographystyle{abbrv-doi-hyperref}
\bibliography{main.bib}

\vfill

\end{document}